\documentclass[journal]{IEEEtran}

\usepackage[T1]{fontenc}
\usepackage{graphicx}
\usepackage{xspace}
\usepackage{cite}
\usepackage{amsthm}
\usepackage{balance}

\usepackage{amsmath,amssymb}
\usepackage{algorithm}
\usepackage{algpseudocode}
\usepackage[cmintegrals]{newtxmath}

\usepackage[utf8]{inputenc}
\usepackage{amsmath}
\newtheorem{example}{Example}
\usepackage{pgfplots}
\pgfplotsset{compat=newest}
\usepackage{tikz}
\usetikzlibrary{plotmarks}
\usetikzlibrary{pgfplots.groupplots}
\usetikzlibrary{external}
\usepackage{xcolor}

\usetikzlibrary{calc}
\newlength{\MyFigureWidth}
\newlength{\MyFigureHeight}
\newcommand{\inputtikz}[1]{
			\tikzsetnextfilename{#1}
			\includegraphics{#1.pdf}
			}
\pgfplotsset{
	every axis legend/.append style={
		legend cell align=left,
		align=left,
		font=\footnotesize
	}
}

\pgfplotsset{
	every axis plot/.append style={
  		line width=1.5pt,
  		mark size=2pt,
      mark options={solid,line width=0.5pt,fill=white!80!.}
  	}
}

\pgfplotsset{
	every axis/.append style={
		label style={font=\footnotesize},
        tick label style={font=\footnotesize}  
    }
}

\definecolor{StrategyIColor}{rgb}{0.00000,0.44700,0.74100}%
\definecolor{StrategyIIColor}{rgb}{1,0,0}%
\definecolor{InterleaverColor}{rgb}{0,0.4,0}%
\definecolor{MPDMMediumColor}{rgb}{0.929,0.694,0.125}%
\definecolor{MPDMLongColor}{rgb}{1,0,0}%

\pgfplotsset{StrategyI/.style={solid,color=StrategyIColor,mark=diamond*}}
\pgfplotsset{StrategyII/.style={solid,color=StrategyIIColor,mark=pentagon*}}
\pgfplotsset{Capacity/.style={solid,line width=2pt,color=CapacityColor}}
\pgfplotsset{CCDMInf/.style={dotted,color=CCDMInfColor}}
\pgfplotsset{Interleaver/.style={solid,color=InterleaverColor, mark=square*}}
\pgfplotsset{MPDMMedium/.style={solid,color=MPDMMediumColor,mark=pentagon*}}
\pgfplotsset{MPDMLong/.style={solid,color=MPDMLongColor,mark=*}}

\definecolor{UniformColor}{rgb}{0.00000,0.44700,0.74100}%
\definecolor{CapacityColor}{rgb}{0,0,0}%
\definecolor{CCDMColor}{rgb}{0,0.6,0}%
\colorlet{SRHCCColor}{blue!60!black}%
\definecolor{MPDMColor}{rgb}{0.929,0.694,0.125}%
\definecolor{MRHCCColor}{rgb}{1,0,0}%
\colorlet{ESSColor}{black}%

\pgfplotsset{Uniform/.style={dashed,color=UniformColor,mark=square*}}
\pgfplotsset{Capacity/.style={solid,line width=2pt,color=CapacityColor}}
\pgfplotsset{CCDM/.style={dotted,color=CCDMColor}}
\pgfplotsset{SRHCC/.style={solid,color=SRHCCColor, mark=diamond*}}
\pgfplotsset{MPDM/.style={solid,color=MPDMColor,mark=pentagon*}}
\pgfplotsset{MRHCC/.style={solid,color=MRHCCColor,mark=*}}
\pgfplotsset{ESS/.style={solid,color=ESSColor,mark=triangle*}}

\newcommand{\MyVec}[1]{\underline{#1}}
\newcommand{\Entr}[1]{\ensuremath{\mathbb{H}\left(#1\right)}}

\usepackage{mathtools}
\DeclarePairedDelimiter\abs{\lvert}{\rvert}%
\let\norm\relax
\DeclarePairedDelimiter\norm{\lVert}{\rVert}%
\let\floor\relax
\DeclarePairedDelimiter\floor{\lfloor}{\rfloor}
\let\ceil\relax
\DeclarePairedDelimiter\ceil{\lceil}{\rceil}
\makeatletter
\let\oldabs\abs
\def\abs{\@ifstar{\oldabs}{\oldabs*}}
\let\oldnorm\norm
\def\norm{\@ifstar{\oldnorm}{\oldnorm*}}
\let\oldfloor\floor
\def\floor{\@ifstar{\oldfloor}{\oldfloor*}}
\let\oldceil\ceil
\def\ceil{\@ifstar{\oldceil}{\oldceil*}}
\makeatother

\newcommand{\MC}[1]{\ensuremath{\text{MC}\!\left( #1 \right)}\xspace}

\newcommand{\Lout}{\ensuremath{n}\xspace}
\newcommand{\Lin}{\ensuremath{k}\xspace}

\newcommand{\Rateloss}{\ensuremath{R_\text{loss}}\xspace}

\newcommand{\Composition}{\ensuremath{\MyVec{C}}\xspace}

\newcommand{\rankrel}{\ensuremath{\MyVec{r}_\text{rel}}\xspace}
\newcommand{\rankrelj}{\ensuremath{r_{\text{rel},j}}\xspace}
\newcommand{\rankrell}{\ensuremath{r_{\text{rel},l}}\xspace}
\newcommand{\rankreljminusone}{\ensuremath{r_{\text{rel},j-1}}\xspace}
\newcommand{\ranktarget}{\ensuremath{r_\text{target}}\xspace}

\newcommand{\nseq}{\ensuremath{n_\text{seq}}\xspace}
\newcommand{\EnergyAvg}{\ensuremath{\mathbb{E}}\xspace}

\newcommand{\rev}[1]{#1}

\usepackage[hidelinks]{hyperref} 

\begin{document}
\title{Huffman-coded Sphere Shaping and Distribution Matching Algorithms via Lookup Tables}

\author{Tobias~Fehenberger,~\IEEEmembership{Member,~IEEE},
        David S. Millar,~\IEEEmembership{Member,~IEEE}, Toshiaki Koike-Akino,~\IEEEmembership{Senior Member,~IEEE}, Keisuke Kojima,~\IEEEmembership{Senior Member,~IEEE}, Kieran Parsons,~\IEEEmembership{Senior Member,~IEEE}, and Helmut Griesser,~\IEEEmembership{Member,~IEEE}
        
\thanks{T.~Fehenberger was with Mitsubishi Electric Research Laboratories. He is now with ADVA, Munich, Germany. E-mail: tfehenberger@adva.com.}
\thanks{D.~S.~Millar, T.~Koike-Akino, K.~Kojima and K.~Parsons are with Mitsubishi Electric Research Laboratories. E-mails: millar@merl.com; koike@merl.com; kojima@merl.com; parsons@merl.com.}%
\thanks{H.~Griesser is with ADVA, Munich, Germany. E-mail: hgriesser@adva.com.}
}

\markboth{T.~Fehenberger \MakeLowercase{\textit{et al.}}: Huffman-coded Sphere Shaping and Distribution Matching Algorithms via Lookup Tables}%
{}

\maketitle

\begin{abstract}
In this paper, we study amplitude shaping schemes for the probabilistic amplitude shaping (PAS) framework as well as algorithms for constant-composition distribution matching (CCDM). Huffman-coded sphere shaping (HCSS) is discussed in detail, which internally uses Huffman coding to determine the composition to be used and relies on conventional CCDM algorithms for mapping and demapping. Numerical simulations show that HCSS closes the performance gap between distribution matching schemes and sphere shaping techniques such as enumerative sphere shaping (ESS). HCSS is based \rev{on} an architecture that is different from the trellis-based setup of ESS. It allows to tailor the used HCSS compositions to the transmission channel and to take into account complexity constraints. We further discuss in detail multiset ranking (MR) and subset ranking (SR) as alternatives to arithmetic-coding (AC) CCDM. The advantage of MR over AC is that it requires less sequential operations for mapping. SR operates on binary alphabets only, which can introduce some additional rate loss when a nonbinary-to-binary transformation is required. However, the binomial coefficients required for SR can be precomputed and stored in a lookup table (LUT). We perform an analysis of rate loss and decoding performance for the proposed techniques and compare them to other prominent amplitude shaping schemes. For medium to long block lengths, MR-HCSS and SR-HCSS are shown \rev{to} have similar performance to ESS. SR-HCSS and uniform 64QAM are compared in additive white Gaussian noise simulations and shaping gains of 0.5~dB and 1~dB are demonstrated with 1~kbit and 100~kbit LUT size, respectively.
\end{abstract}


\IEEEpeerreviewmaketitle

\section{Introduction}
A conventional modulation format whose signal points are equidistant (i.e., lie on a square grid) and equiprobable (i.e., occur with the same probability) can be tailored to the transmission channel for improved performance, such as less required signal-to-noise ratio (SNR) for a fixed throughput. This optimization is known as constellation shaping, and in general two different approaches exist. Geometric shaping, on the one hand, optimizes the position of the constellation points in the complex plane \cite{Sun1993TransIT_GeometricShaping,Fischer2005Book_Shaping,Batshon2010PTL_GeometricShapingIPQ,Lotz2013JLT_GeoShaping}. This approach offers large shaping gains for symbol-wise forward error correction (FEC) decoding. As geometrically shaped constellations often do not have a Gray mapping, it can be challenging to achieve significant shaping gains in bit-interleaved coded modulation (BICM) systems with binary FEC \cite{Millar2018OFC_CodedModulationGS}. Probabilistic shaping (PS), on the other hand, modifies the probability of the constellation symbols that remain on a square grid. Classic PS schemes in the literature include many-to-one mappings \cite{Gallager1968Book_IT}, trellis shaping \cite{Forney1992TransIT_TrellisShaping}, and shell mapping \cite{laroia1994}.

To the best of our knowledge, coded modulation system with PS have been investigated in fiber optics as early as 2012 \cite{Smith2012JLT_CodedModulation,beygi2012adaptive} by using trellis shaping and shell mapping, respectively. More work on PS has been published afterwards, such as \cite{Yankov2014PTL_ProbabilisticShaping,Beygi2014JLT_CodedModulation}. PS attracted a great deal of attention within the fiber-optic community in 2014/15 with the proposal of probabilistic amplitude shaping (PAS) \cite{Boecherer2015TransComm_ProbShaping}. The first demonstrations \cite{Fehenberger2015OFC_ProbShaping, Buchali2015ECOC_ProbShapingExp,Fehenberger2016PTL_MismatchedShaping} were presented shortly after. Ever since then, PAS has been investigated in wireless communications \cite{Gultekin2019Arxiv_ESS,Boecherer2017Arxiv_PDM} and particularly in the fiber-optic literature where various aspects are covered, such as back-to-back sensitivity analysis \cite{Fehenberger2016PTL_MismatchedShaping}, field trials \cite{Cho2017OFC_ShapingFieldTrial}, interplay with digital signal processing \cite{Sasai2019JLT_CPEShaping}, and investigations into fiber nonlinearities, both for infinite-length \cite{Fehenberger2016JLT_ShapingQAM,Renner2017JLT_ShortReachShaping,sillekensSimpleNonlinearityTailoredProbabilistic2018} and finite-length cases \cite{Fehenberger2020JLT_CC,Fehenberger2020OFC_ShapingNLI}.

The success of PAS as coded modulation framework for PS can be attributed to several aspects. As most shaping schemes, it closes the gap to the Shannon limit and offers gains up to 1.53~dB. Furthermore, rate adaptivity is enabled, i.e, the throughput can be varied with fixed modulation order and FEC overhead. Finally, its reverse concatenation principle allows a low-complexity integration of PAS into existing coded modulation system with binary FEC. The main complexity of PAS lies in the block-wise amplitude shaper that transforms a uniform binary input to a sequence of shaped amplitudes at the transmitter. At the receiver, this operation is reversed.

In the initial PAS proposal, a constant-composition distribution matcher (CCDM) \cite{Schulte2016TransIT_DistributionMatcher} is used as amplitude shaper. All shaped output sequences generated by a CCDM have the same empirical distribution, which means that all transmit sequences are permutations of each other. While a CCDM asymptotically achieves ideal performance, its finite-length behavior is suboptimal. The arithmetic coding (AC) algorithm used for CCDM is an inherently sequential method \cite{Schulte2016TransIT_DistributionMatcher}. The combination of long blocks required for low loss and sequential processing has lead to a great deal of research into advanced amplitude shaper schemes that improve upon CCDM. Examples of fixed-length amplitude shapers include multiset-partition distribution matching (MPDM) \cite{Fehenberger2019TCOM_MPDM}, multi-composition DM \cite{Pikus2019Arxiv_MCBLDM}, prefix-free code DM with framing \cite{choPrefixFreeCodeDistribution2019}, enumerative sphere shaping (ESS) \cite{Gultekin2019Arxiv_ESS}, shell mapping (SM) \cite{laroia1994,Schulte2018Arxiv_ShellMapping}, and hierarchical DM \cite{Yoshida2019_HierarchicalDM}.

In the first part of this manuscript \rev{(Sec.~\ref{sec:hcss})}, we study an amplitude shaping technique that uses Huffman coding to address all compositions inside an $n$-dimensional sphere and conventional CCDM algorithms for the mapping or demapping according to the selected composition. This Huffman-coded sphere shaping (HCSS) approach, which has been introduced in \cite{Millar2019ECOC_HCSS}, is a generalization of MPDM and can closely approach the performance of sphere shaping schemes, yet with a different architecture than those techniques. An energy comparison of HCSS with other prominent state-of-the-art shaping schemes is presented to highlight the conceptual difference. We further show numerical simulations results for rate loss and decoding performance. The second part of this manuscript \rev{(Sec.~\ref{sec:ranking_methods})} covers subset and multiset ranking, which are alternatives to the arithmetic coding (AC) method used for CCDM mapping and demapping. Multiset ranking (MR), also briefly mentioned in \cite{Millar2019ECOC_HCSS}, is explained in detail for the first time and algorithms are presented. Subset ranking (SR) has been proposed before in \cite{Fehenberger2020TCOM_PASR} as an alternative implementation for binary-alphabet AC-CCDM that has a greatly reduced number of sequential operations. In this work, an efficient implementation of SR based on a lookup table (LUT) is proposed. We demonstrate that a LUT size of approximately 1~kbit is sufficient to achieve significant shaping gains of 0.5~dB for 64-ary quadrature amplitude modulation (QAM). Increasing the LUT size to 100~kbit can yield shaping gains of up to 1~dB.


\section{Huffman Coded Sphere Shaping (HCSS)}\label{sec:hcss}
In the following, we review fundamentals of PAS and amplitude shapers. We then explain HCSS, and compare it to other shaping algorithms with respect to energy aspects as well as occupation of the shells in an $n$-dimensional sphere.

\subsection{Preliminaries of Amplitude Shapers}
All considered fixed-to-fixed-length amplitude shapers are designed to carry out a seemingly simple mapping operation from a uniformly distributed bit sequence of length $k$ to a shaped sequence of length $n$. This mapping function must be invertible, which makes the considered problem effectively an indexing task. Mathematically speaking, if all length-$n$ sequences are in the target set of the mapping function, we have an injective and non-surjective mapping, which means that each bit input must be mapped onto one shaped sequence, but not every possible sequence of length $n$ is used (as this would just give a uniform distribution on average). Under the assumption typically made in PAS that all errors are corrected by the FEC, the inverse operation, demapping, is then a bijective mapping from shaped sequences to binary data. 

Following \cite{Calderbank1990TransIT_ProbShaping}, we can distinguish between two main classes of amplitude shapers, depending on the properties of the output sequences. We speak of the direct approach if a shaping code is explicitly designed, which corresponds in the nomenclature commonly used in the context of PAS to a distribution matcher, either for a single composition (CCDM) or for various compositions (e.g., MPDM). In the direct approach, the transmission rate follows only indirectly from the \rev{probability mass function (PMF)} and the employed shaper. For the indirect approach, on the other hand, a shaping code is not defined explicitly but obtained by bounding the used signal space by a sphere, where the maximum energy of the sphere determines the transmission rate. Prominent examples of indirect schemes are ESS and SM. More details on this classification can be found in \cite[Sec.~IV]{gultekin2019probabilistic}. The proposed HCSS can be considered a hybrid between these two approaches. As will be explained in Sec.~\ref{ssec:hcss}, a specific HCSS code is designed, but the design premise is to approach a signal space that is bounded in a similar way as for sphere shaping schemes.

\subsection{Rate, Rate Loss, and Related Quantities}
The rate of an amplitude shaper is $k/n$ bits per amplitude symbol. Any finite-length shaping scheme exhibits a rate loss defined as
\begin{equation}
\Rateloss = \Entr{A} - k/n,
\end{equation}
where $\Entr{A}$ is the entropy of the amplitudes $A$ that take on values from the alphabet $\mathcal{A}={a_1,\dots,a_m}$ according to the \rev{PMF} $P_A$. The rate loss directly reduces the throughput that can be achieved with PAS, see \cite[Appendix]{Fehenberger2019TCOM_MPDM}. Hence, for a linear channel such as the additive white Gaussian noise (AWGN) channel, an efficient shaping method seeks to minimize the rate loss for a fixed block length $n$, which typically should also be as small as possible for implementation reasons.\footnote{Notable exceptions are schemes that offer advantages in implementation at the cost of some extra rate loss, such as the parallel-amplitude transformation proposed in \cite{Fehenberger2020TCOM_PASR}.} For a fixed $n$, the number of addressable input bits $k$ is defined as
\begin{equation}\label{eq:k}
k = \floor{\log_2 \nseq},
\end{equation}
where $\floor{\cdot}$ denotes rounding down to the nearest integer and \nseq is the total number of all possible sequences that are output by the respective shaper.

For CCDM, all output sequences have the same composition, i.e., they are permutations of each other. \rev{A composition $\Composition$ is the ordered set of occurrences $ [c_1,\dots,c_m]$ of the amplitudes $[a_1,\dots,a_m]$ in a sequence of length $n=\sum_{i=1}^{m} c_i$. The individual elements of $\Composition$ are formally defined as
\begin{equation}\label{eq:sequence_count}
c_i = \abs{\{\,j: x_j = a_i\,\}}, \quad j\in 1,\ldots,n, \quad i\in 1,\ldots,m.
\end{equation}
}The number of permutations of a composition $\Composition$ is determined by the multinomial coefficient (MC), which is defined as
\begin{equation}\label{eq:MC}
\MC{\Composition} = \frac{n!}{\prod_{i=1}^{m} \left(c_i!\right)}.
\end{equation}
For MPDM, the accumulated permutation count over all permissible compositions must be used. For sphere shaping schemes, the number of used shells is increased until the required \nseq and thus the desired $k$ is achieved. Since the shells in the design process are preferably sorted by energy, using more shells increases the average energy of the scheme. Average energy is defined as
\begin{equation}\label{eq:avg_energy}
\EnergyAvg = \sum_{a \in \mathcal{A}} P_A(a) \cdot a^2.
\end{equation}
For non-constant-composition schemes, the amplitude PMF $P_A$ can be computed from the weighted average of all utilized sequences.

\subsection{Principle of HCSS}\label{ssec:hcss}
\begin{figure}
\begin{center}
\inputtikz{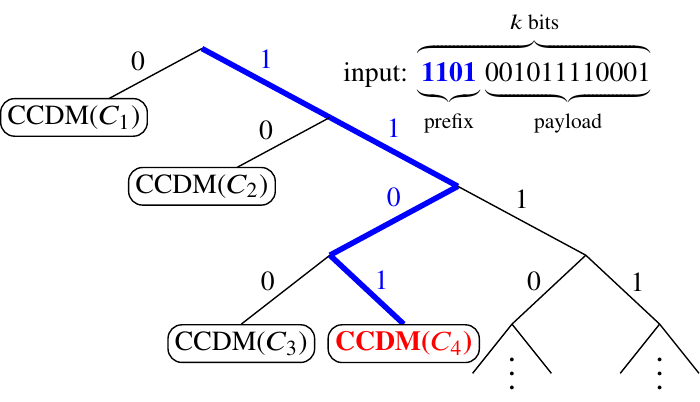}
\end{center}
\caption{Tree structure for Huffman coded sphere shaping (HCSS). A prefix (blue) of the $k$-bit input defines which composition (red) is to be used. The remaining payload is mapped via CCDM methods, such as arithmetic coding or the multiset and subset ranking methods discussed in Sec.~\ref{sec:ranking_methods}.}
\label{fig:tree_structure}
\end{figure}

HCSS is most easily understood when viewed as an extension of CCDM and MPDM. As noted above, CCDM addresses sequences from a single composition and thus, every CCDM output block has an empirical distribution that is the target distribution.\footnote{Note that this is only the case for the one-sided amplitude symbols because the PAS sign bits are not perfectly uniformly distributed, which means that the generated ASK or QAM sequence is not constant composition.} MPDM is an extension of CCDM to multiple compositions with two constraints. Firstly, for pairwise MPDM, only composition pairs are utilized, meaning that for each composition there exists a complement with the same number of permutations such that the target composition, and thus the target PMF, is achieved on average over every pair. Secondly, the number of permutations of each pair must be a power of two to allow construction of a Huffman tree.

HCSS is a further generalization of MPDM where the constraint of having composition pairs is lifted. Instead of just considering pairs, all compositions of length $n$ are sorted by energy and used in order of ascending energy until the total number of permutations allows to address $k$ bits, i.e, until \eqref{eq:k} is fulfilled. For binary input of length $k$, the selection of which composition to be used is based on Huffman coding. Once the composition has been found, the mapping problem is that of a conventional CCDM since we have a constant composition in each node of the binary Huffman tree. In order to construct a Huffman tree, the number of permutations of each composition must be a power of two, which is the same requirement as for MPDM (and trivially also CCDM). As we will see in Sec.~\ref{sec:num_analysis}, introducing structure into the $n$-dimensional sphere results in a small additional rate loss, but allows to use a simple binary Huffman tree and CCDM methods for sphere shaping.

The mapping of a $k$-bit input to a HCSS codeword is illustrated in Fig.~\ref{fig:tree_structure}. First, a prefix (bold blue) determines which composition to use, in this case $C_4$. Mapping the remainder of the input bit sequence to the shaped sequence with composition $C_4$ can be done with any CCDM algorithm, such as arithmetic coding or the methods introduced in Sec.~\ref{sec:ranking_methods}. The binary Huffman tree is constructed by sorting all compositions that fulfill the energy constraint in descending order by their number of permutations, rounding them down to the closest power of two, and assigning a prefix to each of them such that the lengths of the prefix and of the payload are equal to the overall number of addressable input bits $k$.

\begin{figure}
\begin{center}
\scalebox{0.43}{
\inputtikz{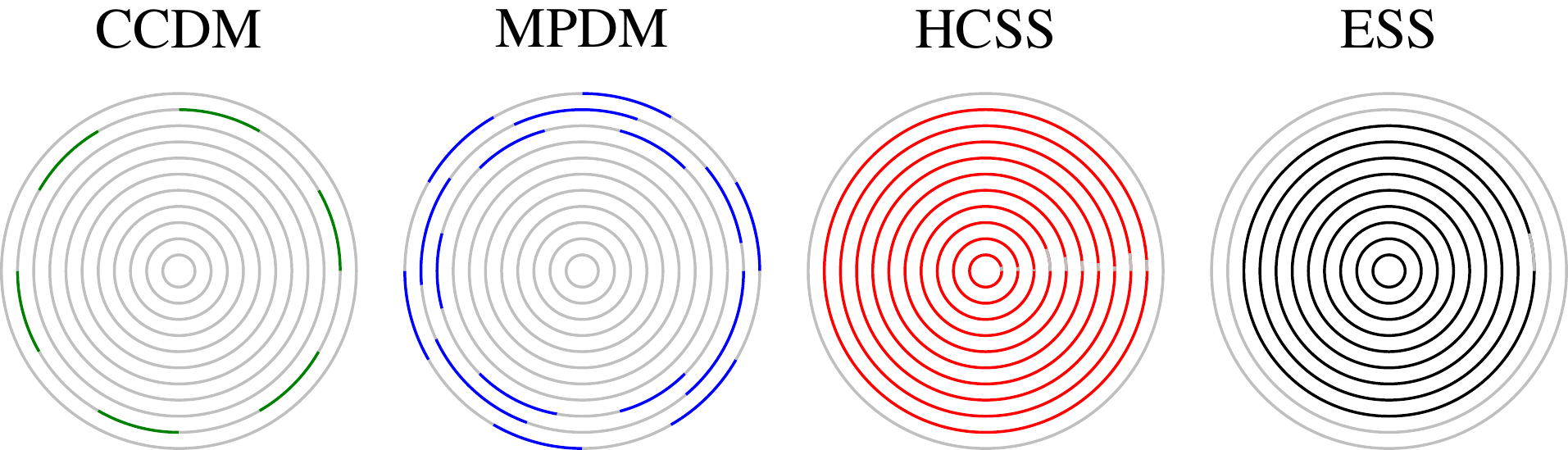}
}
\end{center}
\caption{Illustration of the shell occupation in an $n$-dimensional sphere for various amplitude shapers. Figure extended from \cite[Fig.~6]{gultekin2019probabilistic}.}
\label{fig:Shell2Sphere}
\end{figure}

\subsection{Energy Considerations}
It is well-known that it is beneficial for the AWGN channel to select all possible signal points from within an $n$-dimensional sphere, or equivalently have the signal space bounded by this $n$-sphere \cite{Forney1984JSEL_Shaping,Gultekin2019Arxiv_ESS}. CCDM does not have this property as only sequences of a constant composition \rev{are} used, which corresponds to using only a single fixed-energy shell of this $n$-sphere. Furthermore, as each shell can consist of different compositions, CCDM does in general not fully utilize the shell. This is illustrated in Fig.~\ref{fig:Shell2Sphere}, where each gray circle indicates a shell in the $n$-sphere, and the center of the rings corresponds \rev{to} the zero-energy point. For CCDM, one shell is partly occupied. MPDM combines compositions with their complementary counterpart to achieve the desired distribution on average and thus uses several shells. In contrast to CCDM, each utilized shell can be populated with one or more compositions, which is indicated in Fig.~\ref{fig:Shell2Sphere} by the blue MPDM sections occupying a larger portion of each gray shell. Both HCSS and ESS can use all shells up to a certain maximum energy, and the total number of used output sequences then determines the transmission rate. The difference is that HCSS still operates on compositions, whereas the concept of compositions does not directly exist for ESS. Due to the Huffman tree employed in HCSS, the number of output sequences of each composition must be a power of two, which means that the compositions might not be used fully. This is indicated by the open sections in each HCSS shell in Fig.~\ref{fig:Shell2Sphere}. ESS, in contrast, only has the constraint that the total number of shaped output sequences must be a power of two (a common requirements of all amplitude shaping schemes), which is ideally achieved by not fully using the outermost, highest-energy shell.
Nonetheless, the design principle of HCSS allows to make adjustments to the shaper design that are not feasible with conventional sphere shaping techniques such as ESS or SM, as we will show next.

\subsection{Design Modifications to HCSS}
\begin{figure}
\begin{center}
\inputtikz{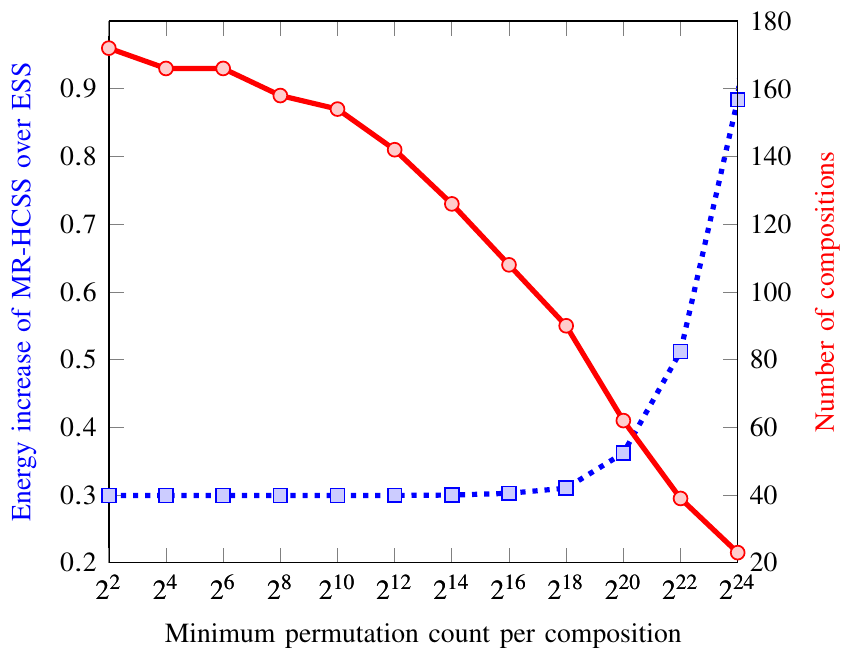}
\end{center}
\caption{Average energy increase of MR-HCSS compared to ESS (left, dotted blue) and Huffman tree depth (right, solid red) versus minimum size of composition in number of permutations for HCSS.}
\label{fig:Energy_vs_MinCompSize}
\end{figure}

The composition-based architecture of HCSS gives the freedom to select which compositions to use and which ones to omit. For instance, compositions can be excluded that might have adverse effects for the nonlinear fiber channel \cite{Fehenberger2019ECOC_NLI_MPDM}. Furthermore, the size of the HCSS Huffman tree can be reduced by omitting composition whose contribution to the total number of permutations is small. This can give a reduction in computational complexity that comes at the expense of an increase in average energy, and thus a small additional rate loss, if less low-energy shells are used. In Fig.~\ref{fig:Energy_vs_MinCompSize}, the energy increase of MR-HCSS over ESS (left axis) and the corresponding Huffman tree depth (right axis) is shown for with $n=20$ and rate 1.5~bit per 1D-symbol (bit/1D-sym). \rev{The energy increase is given in linear scale and is essentially an HCSS penalty, corresponding to the average energy (as defined in \eqref{eq:avg_energy}) of ESS minus that of MR-HCSS. This means that a higher average energy value in Fig.~\ref{fig:Energy_vs_MinCompSize} generally means worse performance.} For computing average energy as defined in \eqref{eq:avg_energy}, the amplitude levels are assumed to be $\mathcal{A}=\{1,3,5,7\}$. The ESS reference has $\EnergyAvg=8.416$\rev{, which is obtained by evaluating the average PMF that is achieved by ESS for the considered parameters.} The horizontal axis shows the minimum number of permutations of each composition, i.e., compositions whose permutation count (given by the multinomial coefficient, see \eqref{eq:MC}) is smaller than that number are not used. We observe from Fig.~\ref{fig:Energy_vs_MinCompSize} that compositions with permutation count up to $2^{18}$ can be excluded from HCSS without a noticeable increase in energy, yet the total number of compositions is significantly reduced from 172 compositions to 90. At this point, the additional rate loss compared to an unmodified HCSS is less than 0.001~bit/1D-sym. When increasing the minimum permutation count beyond $2^{18}$, a penalty arises, and the rate loss increases from 0.098~bit/1D-sym to 0.14~bit/1D-sym. Directly related to the total number of compositions is the maximum Huffman tree depth, which, over the range shown in Fig.~\ref{fig:Energy_vs_MinCompSize}, decreases linearly from 26~bits to 11~bits.

\section{\texorpdfstring{CCDM Algorithms:\\*Subset Ranking and Multiset Ranking}{CCDM Algorithms: Subset Ranking and Multiset Ranking}}\label{sec:ranking_methods}
In this section, we review the subset ranking (SR) and multiset ranking (MR) methods, which are alternatives to arithmetic coding for CCDM. We start with SR, which only supports binary output alphabets, and then cover the more general MR that does not have this restriction. Finally, we discuss the implementation of SR and MR with lookup tables.

\subsection{Subset Ranking}\label{ssec:sr}
Subset ranking is a CCDM algorithm that requires a binary output alphabet. In order to support nonbinary alphabets (i.e., more than two shaped amplitudes, which corresponds to a QAM order higher than 16), a nonbinary-to-binary transformation must be performed, for instance by restricting the PMF to product distributions \cite{Boecherer2017Arxiv_PDM,Pikus2017CommLetters_BLDM} or by using the parallel-amplitude (PA) framework \cite[Sec.~III]{Fehenberger2020TCOM_PASR}, which is used throughout this work for SR. As we will show in Sec.~\ref{sec:num_analysis}, PA introduces some additional rate loss, yet allows higher parallelization than other schemes, see \cite[Sec.~V]{Fehenberger2020TCOM_PASR}.

The idea behind SR is that any binary sequence has an equivalent representation as constant-order subset comprising the indices where one of the two binary symbols occurs in the initial sequence. The cardinality of this subset corresponds to the weight $w$, which is defined as the number of occurrences of one of the two binary symbols. All possible subsets of a certain weight (cardinality) can be sorted according to a specific ordering, such as lexicographical. Each subset in this sorted list of subset can then be ranked, with the rank specifying how many subsets appear in that ordered list prior to the considered subset. With this approach, an invertible mapping function is established between the rank (corresponding to the binary input sequence of length $k$) and the constant-order subset (which is an equivalent representation of the shaped binary output sequence). Therefore, the CCDM mapping operation for a given ranking and with the subset representing the output is referred to as unranking, and demapping is called ranking for which the rank for a certain shaped sequence is to be determined. We will show a simple example illustrating SR in the following. For more details, we refer to \cite[Sec.~IV]{Fehenberger2020TCOM_PASR}.
\rev{
\begin{example}[Subset Ranking]
A binary DM of length $\Lout=10$ and the binary distribution $P_{A}(\alpha)=0.6$, $P_{A}(\beta)=0.4$ can map $\Lin = \floor{\log_2{\dbinom{10}{4}}}=7$~bits to $2^7$ output sequences, each of which contains exactly six times $\alpha$ and four $\beta$-symbols. Now let us represent all possible output sequences by their respective positions of the $\beta$-symbol, where we start counting with one from the left. For instance, $[\beta\alpha\alpha\beta\alpha\alpha\alpha\alpha\beta\beta]$ is equivalent to $[1,4,9,10]$ as $\beta$ occurs at these indices. Next, we sort all possible outputs according to some criterion. In this example, we choose lexicographical order, which means that we start with $[1,2,3,4]$, the next element is $[1,2,3,5]$ and so forth. Suppose the binary word to be mapped is $[1110100]$, which is 116 in the decimal system, and this is referred to as rank. The task of SR mapping is now to choose the sequence in the sorted list that has rank 116, i.e., we pick the element with 116 predecessors, which can be computed (or looked-up) to be $[2,4,7,9]$. This set states the index of $\beta$, giving the final output sequence is $[\alpha\beta\alpha\beta\alpha\alpha\beta\alpha\beta\alpha]$. At the receiver, the inverse operation is carried out by computing the rank based on the received shaped sequence.
\end{example}
}

\subsection{Multiset Ranking}

The previously introduced SR requires a binary alphabet. In the following, we describe multiset ranking (MR) that operates directly on nonbinary alphabets. By lifting the binary-alphabet constraint that can cause a small penalty, MR can be regarded as an implementation alternative to AC-CCDM that, in contrast to SR, does not have extra rate loss and does not require a nonbinary-to-binary transformation.

For the MR method, we distinguish between the target rank, which is simply the input bit sequence to be mapped, and the relative rank, which is defined as the total number of sequences with lower rank, for a given amplitude with a specified lexicographical ordering, occurring in a certain position of the shaped output sequence. Obviously, the relative rank for the smallest amplitude is always zero since the number of preceding sequences for that element by definition must be zero. For the remaining amplitudes, the relative ranks are computed as cumulative sums over the multinomial coefficients of the composition that is updated based on the considered amplitude and output position. For instance, the relative rank of the third amplitude is the sum over the previously computed relative ranks plus the MC of the target composition with the second element reduced by one. The reason for this updating rule for the third amplitude is that we want to determine the number of permutations when the considered position is occupied with the first or second amplitude. When all relative ranks are computed, the amplitude that has the largest relative rank not exceeding the target rank is chosen. Afterwards, the target rank and the remaining composition are updated, and the above steps are carried out until the desired shaped sequence of length $n$ is constructed. The algorithm for MR mapping (unranking) is presented as pseudocode in Algorithm~\ref{alg:MR_mapping}. The inverse operation, MR demapping, is described in Algorithm~\ref{alg:MR_demapping}. For easier notation, we assume that the amplitudes are represented as natural numbers from 1 to $m$, i.e., $\mathcal{A}=\{1,2,\dots,m\}$, and denote the shaped amplitude output sequence as $\MyVec{s}=[s_1,\dots,s_n]$.

\begin{algorithm}[t]
  \caption{Multiset Ranking: Mapping}\label{alg:MR_mapping}
  \begin{algorithmic}[1]
  \Require $\Composition, \ranktarget$ \Comment{composition, target rank}
  \Function{MR\_Mapping}{$\Composition, \ranktarget$}
  	\State $\MyVec{C}_\text{tmp} = \Composition$
    \For{$i$ from 1 to $n$} 
    	\State $\rankrel \gets [0,...,0]$ \Comment{$m$ zeros}
	    \For{$j$ from 2 to $m$} \Comment{all but the first position}
    		\State $\MyVec{t} \gets [0,\dots,0]$ \Comment{temp. variable with $m$ zeros}
    		\State $t_{j-1} \gets 1$ \Comment{1 at index $j-1$}
		    \State $\rankrelj \gets \rankreljminusone + \MC{\MyVec{C}_\text{tmp}-\MyVec{t}}$ 
	    \EndFor
		\State find max. index $l$ where $\ranktarget < \rankrel$
	    \State $s_i \gets a_l$ \Comment{populate output sequence}
	    \State $C_{\text{tmp},l} \gets C_{\text{tmp},l} - 1$ \Comment{update remaining comp.}
	    \State $\ranktarget \gets \ranktarget - \rankrell$ \Comment{update target rank}
    \EndFor
    \State \textbf{return} $\MyVec{s}$ \Comment{return shaped sequence}
  \EndFunction    
  \end{algorithmic}
\end{algorithm}

\begin{algorithm}[t]
  \caption{Multiset Ranking: Demapping}\label{alg:MR_demapping}
  \begin{algorithmic}[1]
  \Require $\Composition, \MyVec{s}$ \Comment{composition, shaped sequence}
  \Function{MR\_Demapping}{$\Composition, \MyVec{s}$}
  	\State $\MyVec{C}_\text{tmp} = \Composition$
    \For{$i$ from 1 to $n$} 
    	\State $\rankrel \gets [0,...,0]$ \Comment{$m$ zeros}
	    \For{$j$ from 2 to $m$} \Comment{all but the first position}
    		\State $\MyVec{t} \gets [0,\dots,0]$ \Comment{temp. variable with $m$ zeros}
    		\State $t_{j-1} \gets 1$ \Comment{1 at index $j-1$}
		    \State $\rankrelj \gets \rankreljminusone + \MC{\MyVec{C}_\text{tmp}-\MyVec{t}}$ 
	    \EndFor
    	\State find index $l$ of $s_i$ \Comment{index in amplitude list}
	    \State $C_{\text{tmp},l} \gets C_{\text{tmp},l} - 1$ \Comment{update remaining comp.}
	    \State $\ranktarget \gets \ranktarget + \rankrell$ \Comment{update target rank}
    \EndFor
    \State \textbf{return} \ranktarget \Comment{return rank}
  \EndFunction    
  \end{algorithmic}
\end{algorithm}

In contrast to arithmetic coding, which is serial in $k$ for mapping and $n$ for demapping, MR is serial in $n$ for mapping and demapping, and thus requires fewer serial operations without incurring additional rate loss. We note that the inner for-loop in Algorithm~\ref{alg:MR_mapping} is considered as fully parallelizable as the computationally demanding step of calculating the MCs can be executed in parallel.

\subsection{Implementation with Lookup Tables}
As we have seen above, the SR and MR algorithms rely on computing binomial and multinomial coefficients, respectively. In the following, we outline the storage requirements when storing these coefficients in a LUT.

The main advantage of SR over AC are the highly parallel algorithms for ranking and unranking whose computational complexity lies mostly in calculating binomial coefficients. For reasonably short block lengths, all required binomial coefficients can be precomputed and stored in a LUT. For a given $n$, the required number of binomial coefficients to be computed is
\begin{equation}\label{eq:lut_sr_1}
\floor{\frac{n}{2}}-1
\end{equation}
as they are symmetric around $n/2$ (i.e., ${n \choose w}\equiv {n \choose n-w}$ for $n>w$) and since ${n\choose 1}=n$. The number of required LUT entries for all DMs with length of at most $n$ is thus
\begin{equation}\label{eq:lut_sr_2}
\sum_{i=4}^{n}\left(\floor{\frac{i}{2}}-1\right),
\end{equation}
where the sum starts at 4 as trivial cases can be omitted which result in a binomial coefficient equal to either 1 or $n$. The size of each LUT entry is
\begin{equation}\label{eq:lut_sr_3}
\ceil{\log_2 {n \choose w}}
\end{equation}
bits, with the maximum size occurring for $w=\floor{n/2}$. With \eqref{eq:lut_sr_1}--\eqref{eq:lut_sr_3} we get an overall LUT size of
\begin{equation}\label{eq:lut_size_sr}
\sum_{i=4}^{n} \sum_{w=2}^{\floor{\frac{i}{2}}} \ceil{\log_2 {i \choose w}}
\end{equation}
bits. As an example, a LUT for all DMs up to length $n=50$ has 14.3~kbit size, with the largest LUT entry requiring 47~bit. A more detailed study of LUT sizes is given in Sec.~\ref{ssec:results_lut}.

For MR of length-$n$ shaping with $m$ amplitudes, the multinomial coefficients of all compositions up to length $n-1$ must be precomputed. The storage requirements can be reduced by taking into account that MCs do not change when the order of the composition elements are permuted, see \eqref{eq:MC}. The overall size of the LUT for MR of length $n$ is thus
\begin{equation}\label{eq:lut_size_mr}
\sum_{i=1}^{n-1} \ceil{\log_2\left(\MC{\mathcal{C}_\text{all}^i}\right)},
\end{equation}
where $\mathcal{C}_\text{all}^i$ denotes the set of all compositions with $m$ elements and block length $i$.

\begin{figure}
\begin{center}
\inputtikz{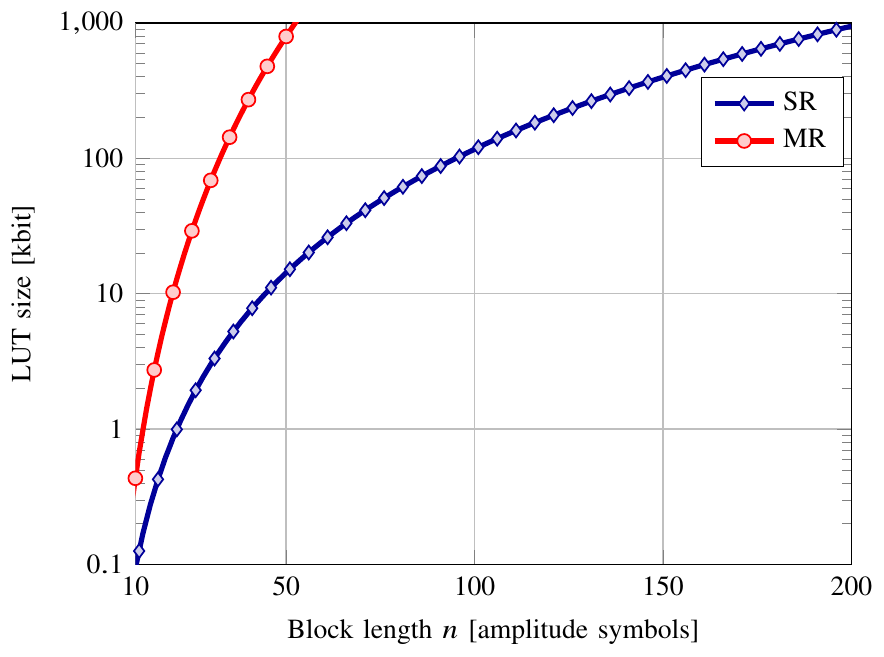}
\end{center}
\caption{LUT size in~kbit versus block length $n$. For MR, $m=4$ amplitudes are assumed.}
\label{fig:LUT_Size_vs_n}
\end{figure}

The required LUT sizes as a function of block length $n$ is shown in Fig.~\ref{fig:LUT_Size_vs_n} for SR and MR. \rev{Note that the same LUTs can be used for matching at the transmitter and inverse matching at the receiver because the binomial coefficients that are to be computed do not change.} We observe that for MR with $m=4$ amplitudes, the LUT size already exceeds 1~Mbit at slightly above $n=50$, which is due to the large number of different MCs that must be stored. For SR, in contrast, only 116~kbit of storage are required for $n=100$, and with 1~kbit of size, a block length of $n=20$ is feasible. We note that the entries of the LUT cannot be used directly for CCDM mapping or demapping as they need to \rev{be} further processed, i.e., added and compared to other numbers such as the rank, see for example Algorithm~\ref{alg:MR_mapping}. The computationally most intense task of computing those coefficients is however greatly alleviated. A numerical analysis of required LUT size for certain shaping gains is presented in Sec.~\ref{ssec:results_lut}.

\section{Numerical Analysis}\label{sec:num_analysis}
In the following, we numerically compare the rate loss, the decoding performance using low-density parity-check (LDPC) codes and the LUT storage requirements for different amplitude shaping methods.

\subsection{Rate Loss}\label{ssec:num_rateloss}
\begin{figure}
\begin{center}
\inputtikz{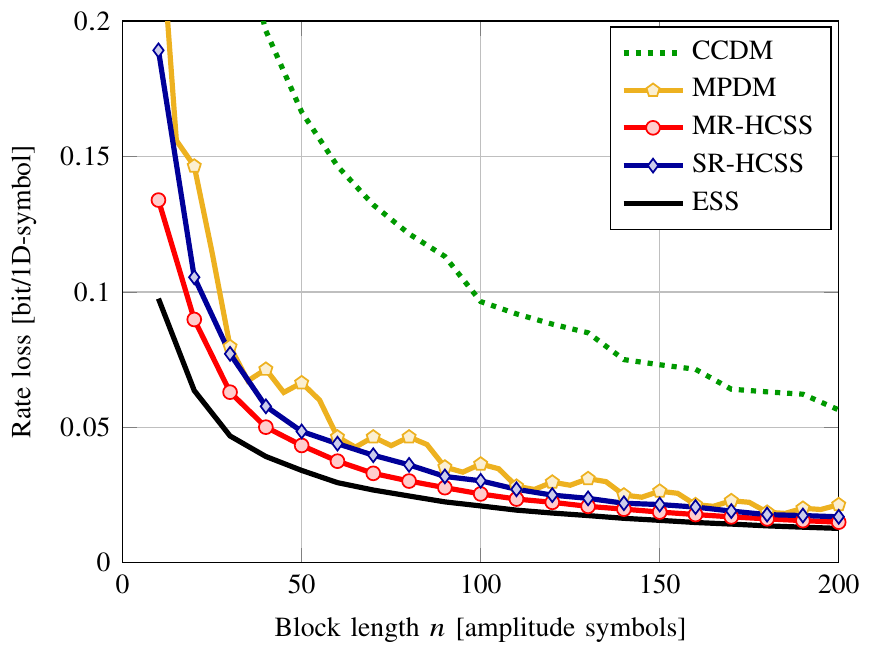}
\end{center}
\caption{Rate loss in bits per 1D amplitude symbol versus $n$. For the two DM schemes, the amplitude distribution is [0.4, 0.3, 0.2, 0.1]. The sphere shaping schemes operate at the same rate as MPDM for each block length.}
\label{fig:RateLoss}
\end{figure}

Figure~\ref{fig:RateLoss} shows the rate loss in bits per 1D amplitude symbol (bit/1D-sym) versus $n$. The amplitude distribution for CCDM and MPDM is $P_A=[0.4, 0.3, 0.2, 0.1]$, and the sphere shaping methods (both HCSS variants and ESS) are set to operate for each $n$ at the rate that MPDM achieves. ESS has the lowest rate loss of all schemes because it uses the signal space in the most energy-efficient way. The HCSS methods have a slightly larger rate loss, and we observe that the gap to ESS reduces for increasing $n$. The reason for this behavior is that at a fixed small $n$, the reduction in the number of input bits $k$ of HCSS compared to ESS translates into a larger absolute rate loss. The performance of MPDM (always using MR internally) is in most cases worse than SR-HCSS, yet MPDM allows to obtain a specific average distribution, which is not feasible with the sphere shaping schemes. Also note that MR-MPDM and MR-HCSS have similar complexity as they differ only in which compositions are used, but the underlying structure of having a Huffman tree and using MR for CCDM operation is identical. The jagged behavior of MPDM is because the employed pairwise partitioning can, depending on $n$, give a different number of composition pairs with a varying permutation count, and thus lead to different rate losses. Interestingly, MPDM sometimes has the same performance as SR-HCSS although the rate losses stem from two different effects. For MPDM, the overall rate loss is due to the suboptimal use of the signal space as only considering pairwise partitions are used. For SR-HCSS, on the other hand, all compositions up to a certain maximum energy are in principle considered, yet the required parallel-amplitude transformation (see Sec.~\ref{ssec:sr}) introduces some penalty. Finally, CCDM has by far the worst rate loss for the considered small-to-medium block lengths.

\begin{figure}
\begin{center}
\inputtikz{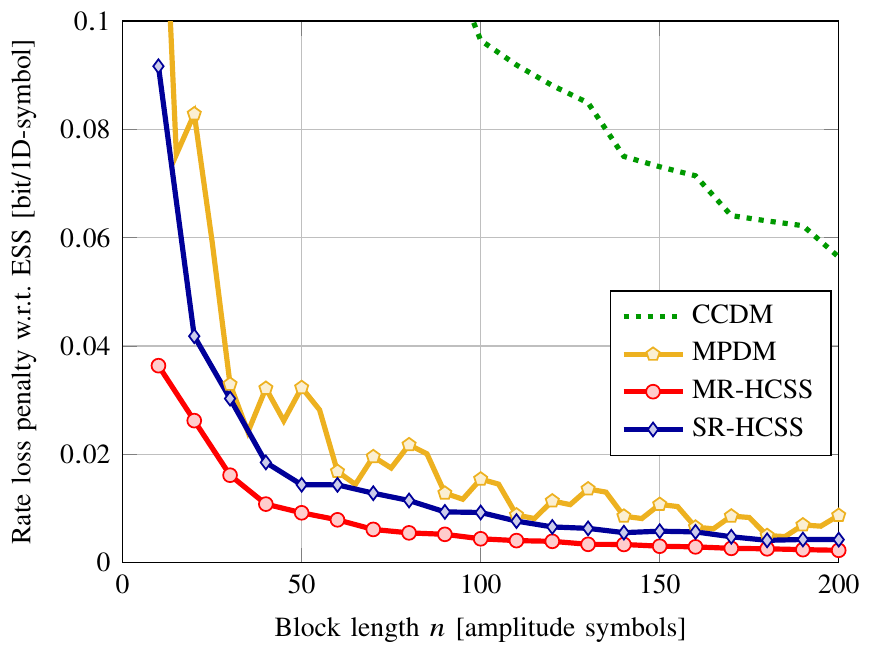}
\end{center}
\caption{Rate loss penalty of various shaping schemes compared to ESS versus block length $n$.}
\label{fig:RateLossPenalty}
\end{figure}

In Fig.~\ref{fig:RateLossPenalty}, the rate loss penalty with respect to ESS is shown. We observe that the penalty for the HCSS schemes is only significant at very short block lengths. Already at $n=50$, the penalty is less than 0.01~bit/1D-sym for MR-HCSS and less than 0.015~bit/1D-sym for SR-HCSS. We can clearly see from Fig.~\ref{fig:RateLossPenalty} how the architecture properties of the HCSS schemes, which are included to facilitate implementation, lead to additional rate loss. First, the power-of-two constraint on the number of permutations of each composition gives some extra rate loss for MR-HCSS. The parallel-amplitude framework used for SR-HCSS results in another penalty, yet allows an efficient LUT-based implementation and highly parallel operation.

\subsection{Decoding Performance}
\begin{figure}
\begin{center}
\inputtikz{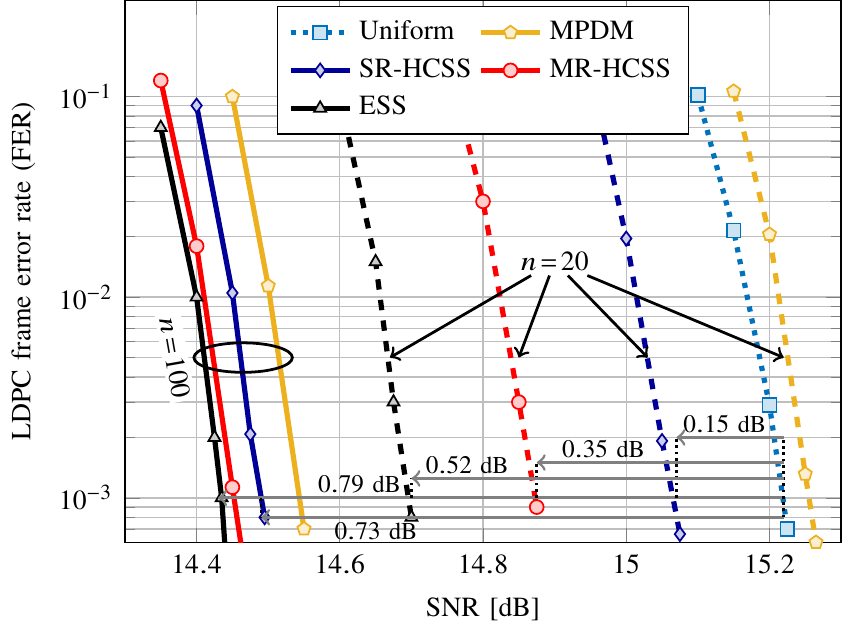}
\end{center}
\caption{Frame error rate (FER) after LDPC decoding versus SNR in dB for the AWGN channel and 64QAM. The information rate is 4.5~bit/2D-sym. DVB-S2 LDPC codes of length 64800 bits are used. All SNR differences are in comparison to uniform signaling and evaluated at a FER of 1e-3.}
\label{fig:FER_SNR}
\end{figure}

For the previously investigated shaping schemes, the frame error rate (FER) after LDPC decoding is shown in Fig.~\ref{fig:FER_SNR}. \rev{Note that we use FER instead of bit error rate (BER) as its shows the percentage of FEC blocks that contain at least one error and thus, must be disregarded or retransmitted. We argue that this metric is more insightful than BER that is not related to the block-based nature of conventional communication systems.} We focus on 64QAM at an information rate of 4.5~bit per 2D symbol (bit/2D-sym) over the AWGN channel and LDPC codes from the DVB-S2 standard with 64000~bit length. \rev{50 decoding iterations were used in all cases}. To achieve 4.5~bit/2D-sym information rate, the code rates are 4/5 for shaped 64QAM and 3/4 for uniform 64QAM. All reported shaping gains are evaluated at a FER of 1e-3. For $n=20$ (dashed curves), a large spread between the different schemes is observed, with SNR savings over uniform 64QAM ranging from $-0.03$~dB (i.e., a penalty because the rate loss is more dominant than the gain from having a shaped QAM distribution) for MPDM, up to 0.52~dB improvement for ESS. The HCSS schemes with MR and SR have shaping gains of 0.15~dB and 0.35~dB, respectively. Overall, the considered schemes differ by more than 0.5~dB of shaping gain. This significant difference in performance is obviously related to the rate loss results discussed in Sec.~\ref{ssec:num_rateloss}. For $n=100$ (solid curves in Fig.~\ref{fig:FER_SNR}), the shaping gains are between 0.73~dB (MPDM) and 0.79~dB (ESS), which is a gain variation of only 0.06~dB. At such a small difference, implementation aspects are believed to be the key differentiator between the different schemes, see e.g. \cite[Sec.~3.3]{Fehenberger2019ECOC_InvitedDM} and \cite[Sec.~VI]{gultekin2019probabilistic}. Finally, we note that the CCDM performance is significantly worse than uniform for $n=20$, and a shaping gain of approximately 0.25~dB is achieved for $n=100$ (both not shown in Fig.~\ref{fig:FER_SNR}).

\subsection{LUT Size versus Shaping Gain}\label{ssec:results_lut}
\begin{figure}
\begin{center}
\inputtikz{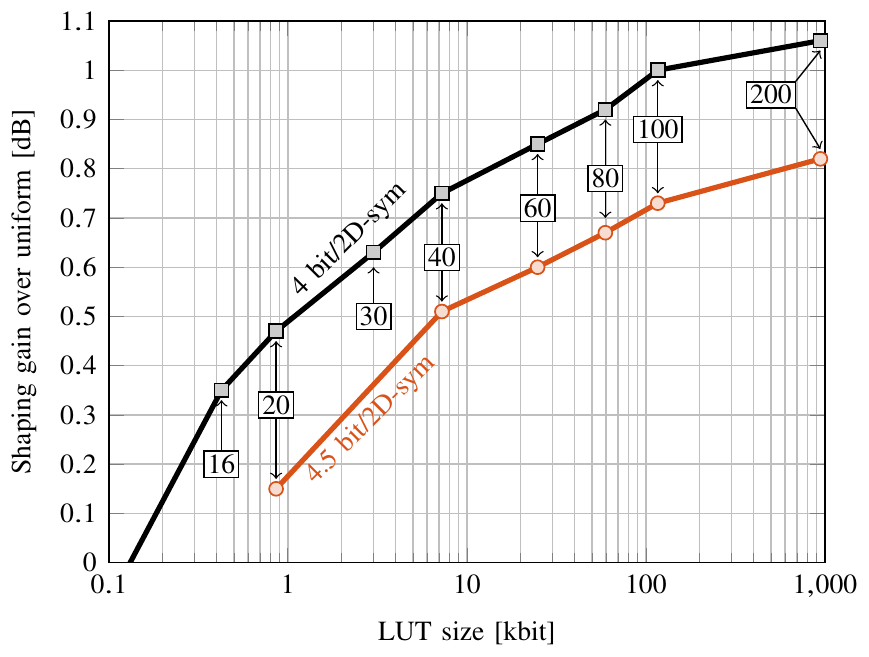}
\end{center}
\caption{Shaping gain over uniform 64QAM with SR-HCSS versus LUT size in~kbit for 4~bit/2D-sym (black, squares) and 4.5~bit/2D-sym (orange, circles). The numbers in squares are the corresponding block lengths $n$.}
\label{fig:Gain_vs_LUT_Size}
\end{figure}

Figure~\ref{fig:Gain_vs_LUT_Size} shows the shaping gain of SR-HCSS for 64QAM versus the size of the LUT required for storing the binomial coefficients. MR is not shown as it has been found to have lower shaping gain than SR for all considered LUT sizes. The two considered information rates of 4~bit/2D-sym and 4.5~bit/2D-sym are obtained with two different configurations, which are 1.5~bit/1D-sym shaping rate and LDPC code rate 5/6 for 4~bit/2D-sym, and 1.85~bit/1D-sym HCSS rate with LDPC code rate 4/5 for 4.5~bit/2D-sym. A similar trend of shaping gain increase with LUT size is observed in Fig.~\ref{fig:Gain_vs_LUT_Size} for both configurations, yet with a difference of more than 0.2~dB between them. \rev{The main reason for the 4~bit/2D-sym case achieving a larger shaping gain is that at this rate, we operate close to AWGN capacity, while the 4.5~bit/2D-sym starts to deviate from capacity due to the limited number of constellation points, see \cite[Fig.~5]{Fehenberger2019TCOM_MPDM}.}
In the following, we focus on the 4~bit/2D-sym case. With less than 1~kbit LUT size (corresponding to $n=20$), a shaping gain of approximately 0.5~dB is achieved. Increasing the LUT to approx. 7~kbit ($n=40$) gives an improvement of 0.75~dB over uniform 64QAM. The next doubling in block length to $n=80$ comes at a LUT size increase by an order of magnitude, yet gives only 0.17~dB extra shaping gain, amounting to an absolute gain of 0.92~dB. Following this trend, the asymptotic (infinite-length) shaping gain of approximately 1.15~dB is achieved to within 0.1~dB with SR-HCSS of length $n=200$, yet at a relatively large LUT of almost 1~Mbit size.

\section{Conclusions}
We have described a sphere shaping scheme that efficiently addresses most of the signal space that lies inside an $n$-dimensional sphere. The proposed Huffman coded sphere shaping (HCSS) is almost as energy-efficient as conventional sphere shaping techniques such as enumerative sphere shaping (ESS). Its construction and mapping methods are, however, highly different from ESS as HCSS uses Huffman coding to determine the composition of the shaped sequence and CCDM methods for mapping and demapping. A numerical analysis of rate loss and decoding performance after LDPC decoding shows that HCSS outperforms all considered DM schemes and performs only slightly worse than ESS for many relevant block lengths, with the performance gap becoming very small for block lengths of $n=50$ and longer. As CCDM algorithms are used for mapping and demapping for the desired composition, HCSS can directly leverage any improvements made in CCDM algorithms.

We have further studied two CCDM mapping/demapping methods that are alternatives to the conventional arithmetic coding (AC). Multiset ranking (MR) mainly relies on computing multinomial coefficients and has a reduced number of sequential operations for mapping than AC, which could facilitate high-throughput applications of probabilistic shaping. We have further studied an implementation of subset ranking (SR) for which the binomial coefficients are precomputed and stored in a lookup table. At the expense of a small additional rate loss compared to MR and AC, SR-HCSS is shown to achieve significant shaping gains over uniform 64QAM of approximately 0.5~dB with LUT sizes of less than 1~kbit. By increasing the LUT size to 100~kbit, the shaping gain increases to approximately 1~dB.

\balance


\bibliographystyle{IEEEtran}


\end{document}